\documentclass[%
 reprint,
 amsmath,amssymb,
 aps,
]{revtex4-2}

\usepackage{graphicx}%
\usepackage{dcolumn}%
\usepackage{bm}%

\usepackage{amssymb}
\usepackage{amsmath}
\usepackage{bm}
\usepackage{xcolor}
\DeclareMathOperator{\adj}{\text{adj}}
\begin{document}

\title{Guiding isotropic active fluids with anisotropic friction}%

\author{Cody D. Schimming}
\affiliation{William H. Miller III Department of Physics \& Astronomy, Johns Hopkins University}

\author{Brian A. Camley}
\affiliation{William H. Miller III Department of Physics \& Astronomy, Johns Hopkins University}
\affiliation{Jenkins Department of Biophysics, Johns Hopkins University}

\begin{abstract}
Inspired by recent experiments of cells accumulating on anisotropic substrates, we study a two-dimensional, compressible, isotropic, active fluid in the presence of anisotropic friction.
We find that regions of anisotropic friction that are patterned as positive topological defects may drive accumulation of an active fluid into a clump, but the robustness of this behavior depends on the initial configuration. %
If the initial azimuthal symmetry is sufficiently broken, we find that patterning asymmetry can instead lead to circular motion of accumulated clumps.
We develop an approximate analytical model to qualitatively explain the motion.
Finally, we use our simplified model to design a substrate pattern that creates directed motion of accumulated clusters along a given path.
\end{abstract}

\maketitle

\section{Introduction}
Active materials are composed of constituents that consume energy on the particle scale resulting in the generation of emergent, macroscopic stresses on the system scale. 
Examples span many length scales and include chemically driven nanoparticles \cite{Perro05,Walther13}, cytoskeletal filament and molecular motor suspensions \cite{Sanchez12,doo18,Kruse24}, cellular tissues \cite{Friedl09,saw18}, and flocks of individual organisms \cite{Makris09,Cavagna14}. 
The variety of these systems leads to a rich diversity in observed behaviors.
For example, active materials can exhibit motility-induced phase separation \cite{Cates15}, low Reynolds number turbulence \cite{wensink12,creppy15,opathalage19}, topological defect proliferation \cite{DeCamp15,sokolov24}, or system-spanning chiral dynamics \cite{Tan22,BZhang22}.

There has been recent interest in the use of external or environmental features to control the behavior of active materials.
This has included the use of boundary geometry \cite{Mahmud09,wioland13,opathalage19,koizumi20,velez24,Velez25}, substrate patterning \cite{turiv20,Endresen21,Kaiyrbekov23,Calderer25,Zhao25}, and external fields \cite{Erglis07,guillamat16,Shim21,Camley18}.
Often, the combination of the symmetry of the active material and the symmetry of the driving factor strongly influence the type of control or behaviors possible. %
For example, external fields may impose a preferred flocking direction, while boundary conditions may induce motile topological defects in active materials with broken orientational symmetry \cite{Kaiser17,opathalage19,Garza23,figueroa22,schimming24,schimming24b,Velez25}. However, symmetry breaking in the environment could potentially drive patterning even if there is no particular broken symmetry in the active material. Simple isotropic contractile active matter models have been successful in characterizing cortical flows in the C. elegans zygote \cite{Gross19,Nishikawa17} and the Xenopus oocyte \cite{Chen24} as well as the formation of periodic patterning in systems of fibroblasts \cite{Palmquist22,camley22}. What happens when an isotropically contractile material is driven by an anisotropic imposed cue? Examples could include contractile cell cortices in the presence of friction gradients \cite{Colin23,Shellard25}, anisotropic micropatterning \cite{Sakamoto24}, or -- in our closest motivating case -- confluent cellular monolayers on substrates with ridgelike patterns \cite{Babakhanova20,turiv20,Endresen21,Kaiyrbekov23,Luo23,Zhao25,Awasthi2025}. 

Here, motivated by these recent experiments, we attempt to understand the effects of substrate anisotropy on isotropic active materials.
We computationally study a model for an isotropic, compressible, active fluid with anisotropic friction.
By studying a simple continuum model, we are able to isolate the effects of isotropic contractility and we show that even a fluid without any topological structure, when paired with friction anisotropy, may accumulate at topological defect centers. %
Further, we show that dynamical states induced by local asymmetry in the friction patterning can arise, showing that local ordering of the material is not necessary for macroscopic displacement.

The rest of the paper is organized as follows.
In Sec.~\ref{sec:Model} we lay out the computational model and show how we incorporate anisotropic friction patterning.
In Sec.~\ref{sec:PhaseTransition} we show that the model exhibits a phase transition in which the homogeneous state is linearly unstable, leading to clustering of the fluid.
Then, in Sec.~\ref{sec:DefectPatterning} we show how anisotropic friction patterning may be used to stabilize clusters in particular locations.
Here, we study the effect of anisotropic friction patterning in the shape of azimuthally symmetric line fields (``$+1$ topological defects'' in the language of nematic liquid crystals) on the stability of the cluster.
We find that geometries of defects ranging from target-shaped patterns with circumferential lines to spiral patterns to radial patterns may all lead to stable accumulation in the center of the defect, provided there is initially an azimuthally symmetric density configuration.
On the other hand, if azimuthal symmetry is initially broken,
the cluster may stabilize in the center of the defect, move continuously around the center of the defect, or be pushed to the edge of the anisotropic region, depending on the geometry of the patterning.
We study this transition numerically, and develop a qualitative analytical model to explain the results.
Finally, in Sec.~\ref{sec:LinearPatterning} we use our intuition from studying the topological defect patterning to posit an anisotropic friction pattern that leads to linear motion of the clusters.

\section{Model} \label{sec:Model}
We study a two-dimensional generalization of the model initially proposed by Bois, J\"{u}licher, and Grill~\cite{Bois11}.
A compressible, isotropic fluid with local density of an active contractile component $c(\mathbf{r})$ (e.g. myosin density, when the model describes a cortex, or cell density when it is applied to a tissue)  and flow velocity $\mathbf{v}({\mathbf{r}})$ satisfies the continuity equation and an overdamped force balance equation (Stokes equation):
\begin{align}
    \frac{\partial c}{\partial t} + \nabla\cdot\mathbf{J} &= 0 \label{eqn:Continuity} \\
     \nabla\cdot \bm{\sigma} + \mathbf{f}_f&= 0 \label{eqn:ForceBalance}
\end{align}
where $\mathbf{f}_f = -\gamma_0 \mathbf{v}$ is the friction with the substrate, with $\gamma_0$ a friction coefficient.
The flux $\mathbf{J}$ is given by advection and Fick's law of diffusion:
\begin{equation}
    \mathbf{J} = c\mathbf{v} -D\nabla c
\end{equation}
where $D$ is the diffusion coefficient of the contractile species. 
The stress tensor $\bm{\sigma}$ is
\begin{equation}
    \bm{\sigma} = \eta\left[\nabla \mathbf{v} + \nabla \mathbf{v}^T - (\nabla\cdot\mathbf{v})\mathbf{I}\right] + \tilde{\eta}(\nabla\cdot\mathbf{v})\mathbf{I} + \bm{\sigma}_a(c)
\end{equation}
where $\eta$ is the shear viscosity, $\tilde{\eta}$ is the bulk viscosity, $\mathbf{I}$ denotes the identity tensor, and $\bm{\sigma}_a$ is an active stress that depends on the local density.

If $\bm{\sigma}_a=0$, the above model represents a typical compressible fluid in the low Reynolds number regime.
The steady state behavior of such a system is a homogeneous density, $c(\mathbf{r}) = c_0$, and zero flow $\mathbf{v}(\mathbf{r}) = 0$.
If $\bm{\sigma}_a \neq 0$, then density variations may generate flows.
Following Ref.~\cite{Bois11}, we assume an active stress of the form
\begin{equation} \label{eqn:ActiveStress}
    \bm{\sigma}_a(c) = \alpha \frac{c}{c^* + c}\mathbf{I}
\end{equation}
where $\alpha$ is the strength of the active stress and $c^*$ is a parameter that determines the density at which the stress saturates.
Equation \eqref{eqn:ActiveStress} represents a contractile stress in the sense that macroscopic forces are generated toward regions of higher density.
That is, if all constituents are ``pulling'' on their neighbors, the regions with higher density will win out. %
The saturating nature of the active stress represents diminishing returns for very large densities, a feature common in many biochemical contexts \cite{Wong18,Marinez-Corral24}.
Variations of the above model have been used to study cellular aggregation in rings of cells \cite{Palmquist22}, actomyosin networks in the cytoskeleton \cite{Ecker21,Staddon22}, and generic activator-inhibitor models \cite{Barberi23,Barberi24}.

\subsection{Anisotropic Friction Patterning}
We now adapt the above model to represent the case where the substrate is anisotropic.
Substrate anisotropy may be induced through several mechanisms, including using an anisotropic medium \cite{Babakhanova20,turiv20,Luo23} or patterning ridges and troughs directly into the surface \cite{Endresen21,Kaiyrbekov23,Zhao25}.
We model this by modifying the force balance equation, Eq.~\eqref{eqn:ForceBalance}, so that the friction between the fluid and substrate 
is given by
\begin{equation} \label{eqn:FrictionForce}
    \mathbf{f}_f=-\gamma_{\parallel} v_{\parallel}\mathbf{\hat{e}}_{\parallel} - \gamma_{\perp}v_{\perp}\mathbf{\hat{e}}_{\perp}
\end{equation}
where $\gamma_{\parallel}$ and $\gamma_{\perp}$ are the friction coefficients in the direction parallel to an easy axis $\mathbf{\hat{e}}_{\parallel}$ and in the direction perpendicular to the easy axis $\mathbf{\hat{e}}_{\perp}$, respectively.

To model a substrate that is patterned, we allow the easy axis to vary in space so that $\mathbf{\hat{e}}_{\parallel}(\mathbf{r}) = g(\mathbf{r})\mathbf{\hat{x}} + h(\mathbf{r})\mathbf{\hat{y}}$ where $g$ and $h$ are the components of $\mathbf{\hat{e}}_{\parallel}$ in the lab frame such that $g^2 + h^2 = 1$.
Further, we will set $\gamma_{\parallel} = \gamma_0$ and $\gamma_{\perp} = (1 + \Delta)\gamma_0$ where $\Delta$ sets the additional friction felt perpendicular to the easy axis.
The limit of isotropic friction is then recovered for $\Delta = 0$.
After locally rotating the friction force density, Eq.~\eqref{eqn:FrictionForce}, to the lab frame, 
the force balance equation, Eq.~\eqref{eqn:ForceBalance}, may be written in terms of a friction coefficient matrix:
\begin{align}
    \gamma_0(\mathbf{I} + \Delta\bm{\Gamma})\mathbf{v} &= \nabla\cdot\bm{\sigma} \label{eqn:AnisoForceBalance} \\
    \bm{\Gamma} &= \begin{pmatrix}
        h^2 & -gh \\
        -gh & g^2
    \end{pmatrix}.
\end{align}

\subsection{Unitless Variables and Numerical Implementation}
To numerically solve Eqs.~\eqref{eqn:Continuity} and \eqref{eqn:AnisoForceBalance} we first rewrite both equations in a non-dimensional form.
We consider a system of size $2L\times2L$ and scale all lengths by $L$ %
and all times by the diffusion time $\tau \equiv L^2/D$ and identify the following parameters
\begin{equation}
    \ell_s^2 \equiv \frac{\eta}{\gamma_0},\quad\ell_b^2 \equiv \frac{\tilde{\eta}}{\gamma_0},\quad \text{Pe} \equiv \frac{\alpha}{\gamma_0 D}
\end{equation}
where $\ell_s$ and $\ell_b$ are hydrodynamic screening lengths corresponding to the shear viscosity and bulk viscosity while $\text{Pe}$ is an effective P\'{e}clet number.
Further, we write the density $c$ in units of $c^*$.
Then, the full dimensionless equations we solve are
\begin{align}
    &\frac{\partial c}{\partial t}+\nabla\cdot(c\mathbf{v}) - \nabla^2c = 0 \label{eqn:CEqn}\\
    &\left(\mathbf{I} + \Delta \bm{\Gamma}\right)\mathbf{v} - \ell_s^2\nabla^2\mathbf{v}-\ell_b^2\nabla(\nabla\cdot\mathbf{v}) -\text{Pe}\frac{\nabla c}{(1+c)^2}= 0 \label{eqn:VEqn}
\end{align}
where all lengths and times are scaled by $L$ and $\tau$ respectively.
There are thus five intrinsic model parameters: $\ell_s$, $\ell_b$, $\text{Pe}$, $\Delta$, and the average density
\begin{equation}
    c_0 = \frac{1}{A}\int c(\mathbf{r},t)\,d\mathbf{r}
\end{equation}
where $A$ is the area of the domain.
Unless otherwise stated, we use the parameter values listed in Table \ref{tbl:Params}.

\begin{table}
\small
  \caption{\ Default parameters used in simulations.}
  \label{tbl:Params}
  \begin{tabular*}{0.48\textwidth}{@{\extracolsep{\fill}}lllll}
    \hline
    $\ell_s$ & $\ell_b$ & $\text{Pe}$ & $\Delta$ & $c_0$ \\
    \hline
    0.075 & 0 & 50 & 3 & 0.329 \\
    \hline
  \end{tabular*}
\end{table}

We solve Eqs.~\eqref{eqn:CEqn} and \eqref{eqn:VEqn} numerically using the MATLAB/C++ finite element package FELICITY \cite{walker18} to discretize in space, and a backwards Euler method (time step $\delta t=0.01$) to discretize in time.
For all computations, we use a square domain such that $x,y\in[-1,1]$ (in dimensionless units as described above) with periodic boundary conditions.
We use periodic boundary conditions since many of the experiments we are inspired by use periodic arrays of substrate patterning \cite{Endresen21,Kaiyrbekov23,Zhao25}. This choice also ensures that the average density $c_0$ remains constant in time.
Additionally, all computations are run until $c$ fails to change in time or for $500$ iterations, i.e. 5 diffusion times $L^2/D$.

\section{Clustering Transition and Linear Stability} \label{sec:PhaseTransition}
For isotropic friction, that is $\Delta = 0$, the model above exhibits a linear instability for specific values of the remaining parameters.
Taking a small perturbation around the ground state, $c(\mathbf{r},t) = c_0 + \delta c(\mathbf{r},t)$ with
\begin{equation}
    \delta c(\mathbf{r},t) = \varepsilon e^{\omega(\mathbf{k})t + i\mathbf{k}\cdot\mathbf{r}}
\end{equation}
where $\varepsilon$ is a small perturbation amplitude, $\omega$ is the perturbation growth rate, and $\mathbf{k}$ is a given wave-vector, a particular solution to Eq.~\eqref{eqn:VEqn} is
\begin{equation} \label{eqn:PerturbationVelocity}
    \mathbf{v}(\mathbf{r},t) = \frac{i \text{Pe}}{(1 + \bar{\ell}^2|\mathbf{k}|^2)(1+c_0)^2}\mathbf{k} \delta c(\mathbf{r},t)
\end{equation}
where $\bar{\ell}^2 = \ell_s^2 + \ell_b^2$.
Substituting Eq.~\eqref{eqn:PerturbationVelocity} into Eq.~\eqref{eqn:CEqn}, keeping terms to linear order in $\varepsilon$, and solving the associated eigenvalue equation yields
\begin{equation} \label{eqn:IsotropicGrowthRate}
    \omega(\mathbf{k}) = -|\mathbf{k}|^2\left[1 - \frac{c_0\text{Pe}}{(1+c_0)^2}\frac{1}{(1 + \bar{\ell}^2|\mathbf{k}|^2)}\right].
\end{equation}
Thus, the homogeneous state $c(\mathbf{r}) = c_0$ is linearly unstable if
\begin{equation}
    \frac{c_0\text{Pe}}{(1 + c_0)^2}\frac{1}{1 + \bar{\ell}^2|\mathbf{k}|^2} > 1.
\end{equation}

\begin{figure}
    \centering
    \includegraphics[width = \columnwidth]{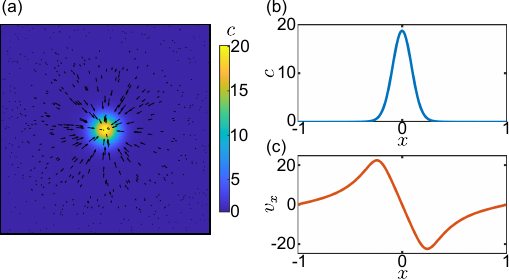}
    \caption{(a) Density field (color) and velocity field (arrows) of a simulation of Eqs.~\eqref{eqn:CEqn} and \eqref{eqn:VEqn} with parameters listed in Table \ref{tbl:Params} and $\Delta = 0$.
    (b) Slice across the $x$-axis of the density field from (a).
    (c) Slice across the $x$-axis of the $x$-component of the velocity field from (a).}
    \label{fig:IsotropicCluster}
\end{figure}

What is the behavior of the system if the instability criterion is met?
We show in Fig.~\ref{fig:IsotropicCluster}(a) a snapshot of a simulation with $\Delta = 0$.
The color in the plot shows the local density $c(\mathbf{r})$ while the arrows show the velocity field.
At long times, the system will develop an accumulation of density in random regions, which will eventually coarsen to a single accumulation spot, or cluster.
Figures \ref{fig:IsotropicCluster}(b,c) show a one dimensional cut of the density and $x$-velocity through a cluster.
This behavior is essentially identical
to the one dimensional system explored in Refs.~\cite{Bois11,Palmquist22}.
The physical explanation for this transition is simple: for a large enough activity, or $\text{Pe}$, the advection generated by the active force is able to overcome the diffusive force dictated by Fick's law.
Hence, the clusters are stabilized by a competition between advection and diffusion. 
This competition sets the cluster size.

The linear stability analysis may be generalized to a uniform anisotropic friction (i.e. a constant matrix $\Gamma$ in Eq.~\eqref{eqn:VEqn}).
To calculate this, it is helpful to work in the $\mathbf{\hat{e}}_{\parallel},\mathbf{\hat{e}}_{\perp}$ basis, or, equivalently, rotating the system so that the easy axis aligns with the $x$-axis.
In this case
\begin{equation*}
    \bm{\Gamma} = \begin{pmatrix}
        0 & 0 \\
        0 & 1
    \end{pmatrix}
\end{equation*}
and we assume, as an ansatz, that $\mathbf{v}$ has the form
\begin{align}
    \mathbf{v}&=(\mathbf{A}\mathbf{k})\delta c \\
    \mathbf{A} &= \begin{pmatrix}
        a_{\parallel} & 0 \\
        0 & a_{\perp}
    \end{pmatrix}.
\end{align}
Substituting the ansatz into Eqs.~\eqref{eqn:VEqn}, solving for $a_{\parallel}$ and $a_{\perp}$, and finding the corresponding growth rate from Eq.~\eqref{eqn:CEqn} yields
\begin{multline} \label{eqn:AnisotropicGrowthRate}
    \omega(\mathbf{k}) = -|\mathbf{k}|^2\Biggl\{1 - \frac{c_0\text{Pe}}{(1+c_0)^2}\\
    \times\frac{1+\frac{\Delta}{1+\ell_s^2|\mathbf{k}|^2}\cos^2\varphi}{1 + \Delta + |\mathbf{k}|^2\left[\ell_s^2+ \ell_b^2\left(1 + \frac{\Delta}{1+\ell_s^2|\mathbf{k}|^2}\cos^2\varphi\right)\right]}\Biggr\}
\end{multline}
where $\varphi$ is the angle between $\mathbf{k}$ and the easy axis.
Note that Eq.~\eqref{eqn:AnisotropicGrowthRate} reduces to Eq.~\eqref{eqn:IsotropicGrowthRate} in the limit $\Delta = 0$, as expected.

\begin{figure}
    \centering
    \includegraphics[width = \columnwidth]{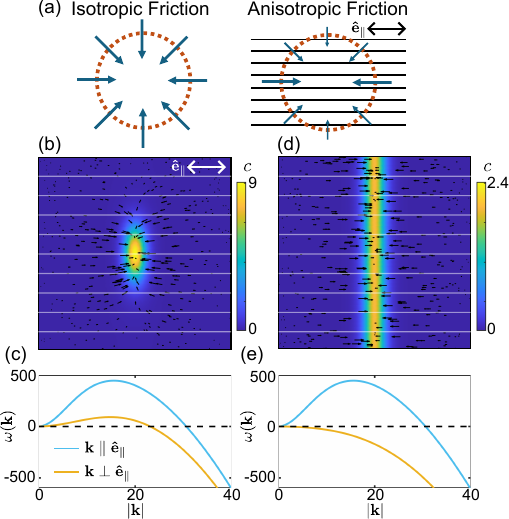}
    \caption{(a) Schematic of the effect of anisotropic friction on the velocity generated by the active force of a circular cluster.
    Isotropic friction yields an isotropic velocity, while anisotropy changes the velocity distribution along the cluster.
    (b) Simulated density field (color) and velocity field (arrows) for parameter values listed in Table \ref{tbl:Params} with constant anisotropic friction, $\mathbf{\hat{e}}_{\parallel}=\mathbf{\hat{x}}$.
    (c) Plots of the linear growth rate, Eq.~\eqref{eqn:AnisotropicGrowthRate}, for the simulation in (b).
    The blue curve shows $\varphi = 0$ while the orange curve shows $\varphi = \pi/2$.
    (d) Simulated density field and velocity field for the same parameters as in (b), except $\Delta = 10$.
    (e) Linear growth rate for the simulation in (d).
    Here $\omega(|\mathbf{k}|,\varphi=\pi/2) < 0$ for all $|\mathbf{k}|$.}
    \label{fig:AnisotropicClustering}
\end{figure}

Unlike in the isotropic model, the growth rate for anisotropic friction now depends on the angle between $\mathbf{k}$ and $\mathbf{\hat{e}}_{\parallel}$.
This can be understood as the active force generated by a gradient in a given direction is subject to a different friction coefficient.
We show this schematically in Fig.~\ref{fig:AnisotropicClustering}(a): Assuming a circular cluster, the velocity induced by the active force would be radially inward and equal magnitude for isotropic friction, but is smaller in directions that are incommensurate with the easy axis of the anisotropic friction. 
The two extreme cases are $\varphi = 0$ ($\mathbf{k}\parallel \mathbf{\hat{e}}_{\parallel}$) and $\varphi = \pi/2$ ($\mathbf{k} \perp \mathbf{\hat{e}}_{\parallel}$).
Because $\varphi = \pi/2$ has the largest friction coefficient, the velocity along this direction will be smallest, and hence this direction provides a lower bound for the growth rate.
Thus, if $\omega(|\mathbf{k}|,\varphi=\pi/2) > 0$ then $\omega(|\mathbf{k}|,\varphi) > 0$ for all $\varphi$ and localized clustering may occur.
We show an example in Fig.~\ref{fig:AnisotropicClustering}(b) of a simulated cluster for the parameters listed in Table \ref{tbl:Params}.
The cluster is anisotropic as the length scales set by advection and diffusion are different in the different directions.
In Fig.~\ref{fig:AnisotropicClustering}(c) we plot the growth rates, Eq.~\eqref{eqn:AnisotropicGrowthRate}, for $\varphi = 0$ and $\varphi = \pi/2$, showing that every direction is unstable for a range of $|\mathbf{k}|$. 

If $\omega(\mathbf{k}) < 0$ for all $|\mathbf{k}|$ and $\varphi = 0$, then the homogeneous state will be stable.
An interesting case occurs when $\omega(\mathbf{k}) > 0$ for $\varphi = 0$, but $\omega(\mathbf{k}) < 0$ for $\varphi = \pi/2$.
Then, the homogeneous state is stable perpendicular to the easy axis, but is unstable along the easy axis, leading to quasi-one dimensional clustering.
We show an example of this in Fig.~\ref{fig:AnisotropicClustering}(d), where a vertical stripe cluster is formed during a simulation with the same parameters as in Fig.~\ref{fig:AnisotropicClustering}(b), except that we increase $\Delta$ to $\Delta = 10$, completely suppressing the instability along the $y$ axis.
In Fig.~\ref{fig:AnisotropicClustering}(e) we plot Eq.~\eqref{eqn:AnisotropicGrowthRate} for $\varphi=0$ and $\varphi=\pi/2$, showing unstable wave-vectors for the case of $\varphi=0$ but $\omega(\mathbf{k}) < 0$ for all $|\mathbf{k}|$ for $\varphi = \pi/2$.

For non-uniform anisotropic friction, it is not analytically tractable to do the same analysis as above.
Nevertheless, it will still be helpful to keep in mind the idea of separate stability along different directions when the anisotropic friction is patterned.
In the following, we work with parameters so that clustering would be stable for uniform anisotropy in both directions $\mathbf{\hat{e}}_{\parallel}$ and $\mathbf{\hat{e}}_{\perp}$.

\section{Topological Defect Patterning} \label{sec:DefectPatterning}
We now study the effect of friction patterning on the clustering of the active fluid.
As mentioned above, we are primarily motivated by recent experiments studying cells on anisotropic substrates \cite{Endresen21,Kaiyrbekov23,Zhao25}.
In those experiments, the primary patterns studied were topological defects, that is, points in a line or vector field which are singular.

\begin{figure}
    \centering
    \includegraphics[width = \columnwidth]{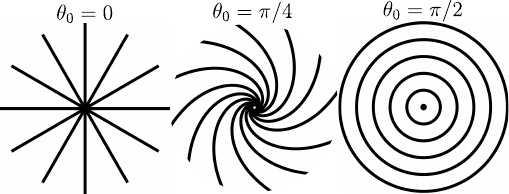}
    \caption{Visualization of easy axis patterns for topological defects with $\theta_0 = 0, \pi/4,$ and $\pi/2$.}
    \label{fig:DefectPatterns}
\end{figure}

Here, our goal is to model the effect of anisotropic substrate patterning on an isotropic active fluid.
We focus on anisotropic friction patterning that mimics a $+1$ winding topological defect so that the easy axis is
\begin{equation}
    \mathbf{\hat{e}}_{\parallel} = \cos(\theta+\theta_0)\mathbf{\hat{x}} +\sin(\theta+\theta_0)\mathbf{\hat{y}} \label{eq:epar_defect}
\end{equation}
where $\theta$ is the azimuthal angle in polar coordinates and $\theta_0$ is an offset angle that determines the shape of the topological defect.
Figure \ref{fig:DefectPatterns} shows the easy axis patterns for $\theta_0 = 0$, $\pi/4$, and $\pi/2$.
Note that any pattern with $0 < \theta_0 < \pi/2$ will have a spiral shape and breaks chiral symmetry.

In what follows, all topological defect patterns occupy a circle of radius $R = 0.5$, above which we set $\Delta = 0$.
We do this to avoid boundary effects where clusters may overlap anisotropic regions across the periodic boundary.

\begin{figure*}
    \centering
    \includegraphics[width = \textwidth]{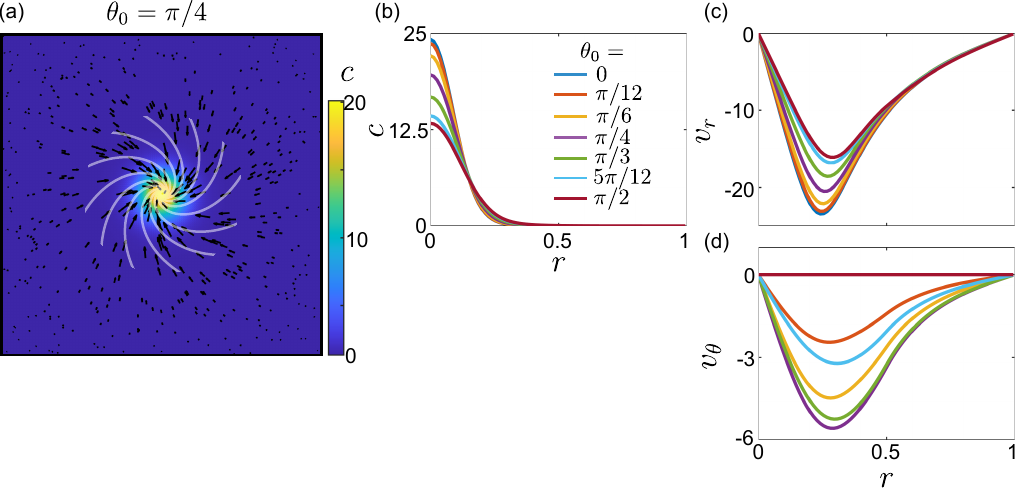}
    \caption{(a) Simulated steady state density field (color) and velocity field (arrows) with anisotropic friction patterned as a topological defect with $\theta_0 = \pi/4$.
    (b) Radial cut of steady state density fields for simulations with varying $\theta_0$.
    (c) Radial cut of the radial component of the steady state velocity field for simulations with varying $\theta_0$.
    (d) Radial cut of the azimuthal component of the steady state velocity field for simulations with varying $\theta_0$.}
    \label{fig:StableCenter}
\end{figure*}

\subsection{Defect Center Accumulation}

We first simulate the case in which the density field is initially slightly larger in the center of the domain, peaked at the topological defect core in the anisotropic friction pattern.
Specifically, we initialize the system with
\begin{equation} \label{eqn:InitSymmetricBump}
    c(r,0) = e^{-\frac{r^2}{0.1}} + \bar{c},
\end{equation}
where $\bar{c} = 0.25$ is a constant background density.
This initial condition is similar to the condition at the time of confluence reported in the experiments of Ref.~\cite{Zhao25}.

We find, for all values of $\theta_0$, that stable clusters may form in the center of the topological defect patterning.
Figure \ref{fig:StableCenter}(a) shows a characteristic two-dimensional snapshot of a configuration in steady state with $\theta_0 = \pi/4$.
In this case the velocity field has a nonzero azimuthal component induced by the spirality of the anisotropic friction.
The size of the steady state cluster at the center of the defect patterning is larger for larger $\theta_0$.
We show this in Fig.~\ref{fig:StableCenter}(b), which plots radial cuts of the steady state density field for simulations with varying $\theta_0$. %
As $\theta_0$ increases, the maximum density decreases and the spatial extent of the cluster increases.
This can be understood as a result of the azimuthal symmetry of the configuration:
As $\theta_0$ increases, the friction in in the radial direction increases, reducing the magnitude of the flow and, hence, the amount of advection.
The radial component of the velocity field is plotted in Fig.~\ref{fig:StableCenter}(c) for the same range of $\theta_0$, showing that the flow magnitude is smallest for $\theta_0 = \pi/2$.
We also plot the azimuthal component of the velocity field in Fig.~\ref{fig:StableCenter}(d) for the same simulations, showing that azimuthal symmetry is broken for $\theta_0\neq 0,\pi/2$.

Our qualitative picture of accumulation is not strongly dependent on the system parameters. For instance, changing $\ell_s$ and $\ell_b$ changes the size of cluster, but as long as the cluster size is smaller than the region of anisotropy, the presence of central accumulation, radial inward flow, and azimuthal flow once a threshold P\'eclet number is reached is unchanged.
If the screening lengths (and hence, the size of the clusters) approach the size of the region of anisotropy $R$ (which is not far from the system size), we found that while we see initial accumulation at the center of the defect patterns with $\theta_0 \approx\pi/2$, this behavior has a longer-term instability, as the clusters interact across the periodic boundaries and tend to move to the region between anisotropy regions at the system boundary.
A larger value of $\Delta$ increases the azimuthal component of the velocity, as well as changing the relative cluster size for different values of $\theta_0$, but again does not qualitatively change the behavior 
of the system, except to require a larger value of $\text{Pe}$ to transition to clustering for $\theta_0 \approx \pi/2$, which is intuitive given our results for the linear stability in constant anisotropy in Sec.~\ref{sec:PhaseTransition}.
We plot radial cuts of configurations [similar to Fig.~\ref{fig:StableCenter}(b--c)] for a few examples of different model parameters in a Supplementary Figure\cite{SuppNote25}.

\begin{figure}
    \centering
    \includegraphics[width = \columnwidth]{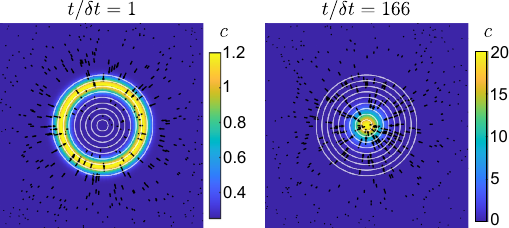}
    \caption{Time snapshots of simulated density field (color) and velocity field (arrows) for anisotropic friction patterned as a topological defect with $\theta_0 = \pi/2$ initialized with a symmetric ring of high density, Eq.~\eqref{eqn:RingDensity}.
    The first snapshot shows the initial condition while the second snapshot shows the final, stable configuration.}
    \label{fig:AnnulusStability}
\end{figure}

Our initial condition has a region of slightly higher density at the origin. 
Given that the active contractile fluid aggregates into clumps even in the absence of friction anisotropy, is the appearance of the aggregation at the origin really arising from the substrate pattern, or is the initial condition just breaking the initial translational symmetry, nucleating an aggregation at the origin?
We find that enhancement of density at the origin in the initial condition is not necessary to have aggregation at the origin. 
Initializing the system with an azimuthally symmetric ring of high density
\begin{equation} \label{eqn:RingDensity}
    c(r,0) = e^{-\frac{(r-0.4)^2}{0.01}} + \bar{c},
\end{equation}
still results in accumulation at the center for all $\theta_0$, as with the initial condition in Eq.~\eqref{eqn:InitSymmetricBump}.
In Fig.~\ref{fig:AnnulusStability} we show snapshots of the initial condition and the final stable state of a simulation with $\theta_0=\pi/2$, and we further show a movie of the simulation in Supplemental Movie 1\cite{SuppNote25}.
If the initial condition is not azimuthally symmetric or has large fluctuations, then we do not observe the accumulation behavior for all angles $\theta_0$, and occasionally see dramatically different behavior including persistent cluster migration, which we discuss in detail in the next subsection.

Stable clustering in the center of a topological defect pattern for all values of $\theta_0$ was also observed experimentally in Ref.~\cite{Zhao25}.
While the cells studied in Ref.~\cite{Zhao25} are elongated, and tend to show nematic ordering along the ridge direction, the model we study does not take into account any anisotropy of constituents.
That our results are qualitatively similar shows that aggregation alone does not imply active nematic behavior, and that the density increases of Ref. \cite{Zhao25} -- though not any of the ordering dynamics -- could be created by alternate patterning mechanisms that do not strongly depend on the ordering of cells.  %

\subsection{Circular Cluster Motion}
Above, we found that initial increases of density either at the pattern center or azimuthally symmetric about it lead to aggregation at the center. What if we place an initial clump away from the center -- does it simply become attracted to the center of the defect pattern? 
We move the initial density peak along the $x$-axis:
\begin{equation}
    c(x,y,0) = e^{-\frac{1}{0.1}\left[(x+0.25)^2+y^2\right]} + \bar{c}.
\end{equation}

\begin{figure}
    \centering
    \includegraphics[width = \columnwidth]{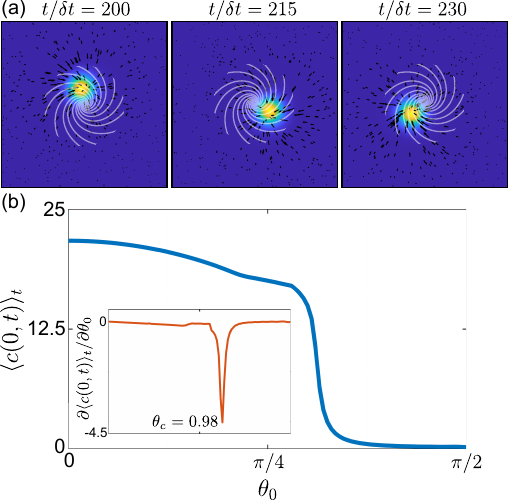}
    \caption{(a) Time snapshots of the simulated density field (color) and velocity field (arrows) for anisotropic friction patterned as a topological defect with $\theta_0 = \pi/3$.
    (b) Average density at the center of the defect pattern, $\langle c(0,t)\rangle_t$, as a function of $\theta_0$.
    Inset: $\partial\langle c(0,t)\rangle_t/\partial\theta_0$ versus $\theta_0$.}
    \label{fig:CircularClusterMotion}
\end{figure}

For $\theta_0 \leq \pi/4$, the system still reaches a steady state with an accumulation at the defect core.
Even if the cluster initially materializes outside the core, it is still eventually guided to the center by the friction patterning (see e.g. Supplemental Movie 2\cite{SuppNote25}).
On the other hand, for some critical angle $\theta_c > \pi/4$, the system transitions to a state where a cluster still forms, but moves radially to the edge of the anisotropic region.
If $\theta_c < \theta_0 <\pi/2$, the cluster will move around the center of the defect in a direction given by the azimuthal symmetry breaking of the anisotropy and will not reach a steady state.
In Fig.~\ref{fig:CircularClusterMotion}(a) we plot several time snapshots of the density field for a simulation with $\theta_0 = \pi/3$, showing that the cluster moves around the center of the defect pattern (see also Supplemental Movie 3\cite{SuppNote25}).
For $\theta_0 = \pi/2$, the cluster will move to the edge of the anisotropic region, but because chiral symmetry is restored to the friction pattern, the cluster will reach a steady state and not move.

To quantitatively characterize this transition between central accumulation and outward motion, we plot in Fig.~\ref{fig:CircularClusterMotion}(b) the time average of the density at the center of the defect, $\langle c(0,t)\rangle_t$ as a function of $\theta_0$.
If the patterning induces central accumulation, $\langle c(0,t)\rangle_t > 0$ since the cluster quickly moves to the center of the pattern.
On the other hand, if the cluster moves away from and around the center, then $\langle c(0,t)\rangle_t \approx 0$.
The transition to $\langle c(0,t)\rangle_t$ characterizes $\theta_c$.
Indeed, in the inset of Fig.~\ref{fig:CircularClusterMotion}(b) we plot $\partial\langle c(0,t)\rangle/\partial\theta_0$, which shows a a strong negative peak at an isolated value of $\theta_0$.
Taking the location of this peak to define the critical angle, we find $\theta_c \approx 0.98$.

Fig. \ref{fig:CircularClusterMotion} shows that slight changes to our system geometry let us switch between concentration accumulating in a single targeted position and persistent migration. Both accumulation and persistent motion may be useful in designing substrates with well defined regions of accumulation, allowing for longer-range mass transport. Unlike active polar or nematic fluids, where intrinsic symmetry breaking leads to sustained motion, such as in flocks or active nematics \cite{Vicsek95,marchetti13,doo18}, here, the fluid itself is isotropic, and, instead, the environment breaks the symmetry and induces motion. The lack of spontaneous symmetry breaking is, in some sense, an advantage of this scheme in terms of control of directionality. For instance, for the directed motion we show in Fig. \ref{fig:CircularClusterMotion}, even if the friction anisotropy were very weak, we would not ever expect a spontaneous rotation in the opposite sense of the guiding cue, but only an absence of rotation. By contrast, in the presence of spontaneous symmetry breaking in a confined active polar fluid\cite{Doxzen2013}, as the guiding cue is weakened, we would expect an equal mix of clockwise and counter-clockwise rotation.  %

\subsection{Cluster Motion Model} \label{subsec:ApproximateModel}
When would we expect to create persistent migration of an active matter clump? What sets the transition between localization to the defect center and persistent migration? We can qualitatively understand our numerical results of the previous section with a simple analytical model for the motion of an accumulated cluster in the presence of anisotropic friction.
We first note that the center of mass of a configuration is given by
\begin{equation}
    \mathbf{X}_c(t) = \frac{1}{M}\int\mathbf{r} c(\mathbf{r},t) \, d\mathbf{r}.
\end{equation}
where $M = \int c\,d\mathbf{r}$ is the total mass.
The center of mass velocity is then 
\begin{equation} \label{eqn:CenterMassVelocity}
    \mathbf{V}_c(t) = \frac{\partial\mathbf{X}_c}{\partial t}=\frac{1}{M}\int\mathbf{r}\frac{\partial c}{\partial t}\, d\mathbf{r}= \frac{1}{M}\int\mathbf{J}(\mathbf{r},t)\,d\mathbf{r}
\end{equation}
where $\mathbf{J} = c\mathbf{v} -\nabla c$ is the flux.
The last equality uses the continuity equation, Eq.~\eqref{eqn:Continuity}, and integration by parts assuming an infinite system.
We can thus predict the instantaneous center of mass velocity by computing $\mathbf{J}$, which is only a function of $c$ and its spatial derivatives, and integrating.

To obtain an approximation of $\mathbf{J}$ we make several approximations of the full model.
We first assume that a cluster is given by a circularly symmetric Gaussian profile
\begin{equation} \label{eqn:CApprox}
    \tilde{c}(\mathbf{r}) = \frac{c_0}{\pi a^2}\exp\left[-\frac{1}{a^2}\left((x-x_0)^2+(y-y_0)^2\right)\right]
\end{equation}
where $a$ is a parameter that quantifies the size of the cluster and $x_0$ and $y_0$ give the position of the cluster center.
The Gaussian ansatz is motivated by the shape of clusters in the isotropic friction limit, but will only be an approximation to the true profile $c(\mathbf{r})$.
We work in the dry limit of the model so that $\ell_s=\ell_b=0$ in Eq.~\eqref{eqn:VEqn}.
We also assume the active force in Eq.~\eqref{eqn:VEqn} is given by $f_a = \text{Pe}\nabla c$ instead of the saturating form used in the simulations.
Finally, we assume the system and region of anisotropy is infinite.

After all of these approximations, the problem becomes analytically tractable, with the force balance equation becoming
\begin{equation}
    \left(\mathbf{I} + \Delta \bm{\Gamma}\right)\mathbf{v} = \text{Pe}\nabla \tilde{c}.
\end{equation}
Inverting the friction coefficient matrix, the velocity field is then
\begin{equation} \label{eqn:ApproxVelocity}
    \mathbf{v} = \frac{\text{Pe}}{1+\Delta}\left[\mathbf{I}+\Delta \tilde{\bm{\Gamma}} \right]\nabla \tilde{c}
\end{equation}
where $\tilde{\bm{\Gamma}} = \adj(\bm{\Gamma})$ is the adjugate of $\bm{\Gamma}$.
The flux (in dimensionless units as defined in Sec.~\ref{sec:Model}) is then 
\begin{equation} 
    \mathbf{J} = \left[\left(\frac{\tilde{c}\text{Pe}}{1+\Delta} - 1\right)\mathbf{I} + \frac{\Delta \tilde{c}  \text{Pe}}{1+\Delta}\tilde{\bm{\Gamma}}\right]\nabla \tilde{c}. \label{eq:flux_approximation}
\end{equation}
For our approximate cluster configuration, $\int \tilde{c}^n\nabla \tilde{c}\,d\mathbf{r}=0$ for any power $n$. 
For this reason, we will often split $\mathbf{J} = \mathbf{J}^{\text{iso}} + \mathbf{J}^{\text{ani}}$, where $\mathbf{J}^{\text{iso}}$ is the isotropic component of the flux, proportional to $\mathbf{I}\nabla\tilde{c}$ in Eq.~\eqref{eq:flux_approximation}, and $\mathbf{J}^{\text{ani}}$ is the anisotropic component of the flux, proportional to $\tilde{\bm{\Gamma}}\nabla\tilde{c}$.
If $\Delta$ does not vary in space, the isotropic component of $\mathbf{J}$ will integrate to zero and not contribute to $\mathbf{V}_c$.
Similarly, the center of mass velocity will be $\mathbf{V}_c = 0$ for any constant anisotropy matrix $\bm{\Gamma}$. 
This is borne out in our simulations with a constant anisotropy (see Fig.~\ref{fig:AnisotropicClustering}); while the shape of the cluster (higher mass moments) may change, its center of mass position does not change.
On the other hand, if the friction anisotropy is patterned so that $\tilde{\bm{\Gamma}}$ varies in space, only $\mathbf{J}^{\text{ani}}$ will contribute to $\mathbf{V}_c$ (again, assuming $\Delta$ is equal everywhere).
Thus, the center of mass velocity is predicted to scale as $|\mathbf{V}_c| \propto \text{Pe} \Delta/(1 + \Delta)$, which we test numerically in the next section.

If the anisotropy is the topological defect pattern of Eq. \ref{eq:epar_defect}, then
\begin{equation}
    \tilde{\bm{\Gamma}} = \frac{1}{2}\begin{pmatrix}
        1 +\cos2(\theta+\theta_0) & \sin2(\theta + \theta_0) \\
        \sin2(\theta+\theta_0) & 1 - \cos2(\theta+\theta_0)
    \end{pmatrix}.
\end{equation}
If $x_0=y_0=0$ so that the cluster is centered at the core of the defect, the center of mass velocity may be computed:
\begin{equation}
    \mathbf{V}_c = -\frac{2}{c_0 a^2}\int r^2 c(r)^2\left[\tilde{\bm{\Gamma}}\mathbf{\hat{r}}\right]\, drd\theta = 0
\end{equation}
where $\mathbf{\hat{r}}$ is the radial unit vector.
Because of the azimuthal symmetry of the configuration and anisotropy, the cluster velocity is zero, as we observed from the simulations above.

If we allow the cluster to be centered away from the defect center, say at a distance $r_0$, $\tilde{c}$ is no longer azimuthally symmetric.
In this case, rotating our axes so that the cluster center lies on the $x$-axis, the anisotropic part of the flux is
\begin{multline} \label{eqn:DefectAniCurrent}
    \mathbf{J}^\textrm{ani} = -\frac{c_0^2}{\pi^2 a^6}\frac{\Delta\text{Pe}}{1+\Delta}\exp\left[-\frac{2}{a^2}\left(r^2+r_0^2-2rr_0\cos\theta\right)\right] \\ \times\tilde{\bm{\Gamma}}\left[\left(r\cos\theta -r_0\right)\mathbf{\hat{x}} + r\sin\theta\mathbf{\hat{y}}\right].
\end{multline}
Integrating the $x$-component of Eq.~\eqref{eqn:DefectAniCurrent} will yield the radial component of the cluster velocity, while the $y$-component yields the azimuthal component.
We compute a non-vanishing center of mass velocity (the integration is shown in the appendix):
\begin{equation} \label{eqn:DefectClusterVel}
    \mathbf{V}_c = -\frac{c_0}{8\pi a^2}\frac{\Delta \text{Pe}}{1+\Delta}\frac{1 - \exp\left[-\frac{2 r_0^2}{a^2}\right]}{r_0}\left[\cos2\theta_0 \mathbf{\hat{r}} + \sin2\theta_0 \bm{\hat{\theta}}\right]
\end{equation}
where $\bm{\hat{\theta}}$ is the azimuthal unit vector.

Equation \eqref{eqn:DefectClusterVel} makes several predictions about the motion of idealized clusters.
First, the cluster velocity $\mathbf{V}_c \to 0$ when $r_0 \to 0$, coinciding with the prediction for a cluster centered at the defect center.
Additionally, the azimuthal component of the cluster velocity is nonzero and has a consistent sign for $\theta_0 \in (0,\pi/2)$ -- clusters circulate the center of the defect pattern due to the breaking of chiral symmetry.
Finally, the radial component of the cluster velocity is negative for $\theta_0 < \pi/4$ and positive for $\theta_0 > \pi/4$, predicting that there should be a critical angle (namely, $\theta_c = \pi/4$) where the motion transitions from inward toward the center of the defect, to outward away from the center of the defect.
All of these predictions are qualitatively borne out in the simulations presented above: Clusters initialized at the center of the defect pattern are stationary; for $0 <\theta_0 < \pi/2$ clusters initialized away from the center move in the $-\bm{\hat{\theta}}$ direction; and there exists a critical angle $\theta_c$ in which clusters transition between moving toward or away from the center.

\begin{figure}
    \centering
    \includegraphics[width = \columnwidth]{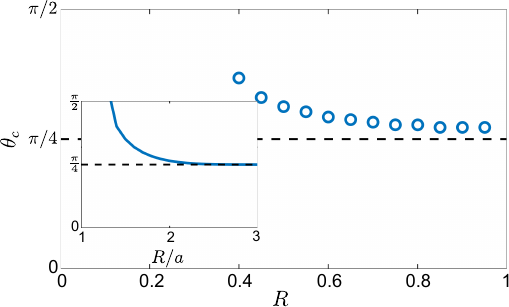}
    \caption{Critical angle of the inward-outward cluster velocity transition, $\theta_c$, as a function of anisotropy region radius, $R$, for simulations of the model.
    Inset: Numerically obtained predicted critical angle from the approximate model of Sec.~\ref{subsec:ApproximateModel}.
    The integrals that were numerically computed are given in Eq.~\eqref{eqn:CriticalAngle} for $r_0/a = 1$. }
    \label{fig:CriticalThetaDistance}
\end{figure}

The predicted $\theta_c = \pi/4 \approx 0.79$ differs from the $\theta_c \approx 0.98$ found in the simulations above.
We performed additional simulations varying $\ell_s$, $\ell_b$, $\Delta$ and $c_0$
and found $\theta_c$ was only very weakly dependent on the parameters as long as they maintained the same linear instability found in Sec.~\ref{sec:PhaseTransition}.
Further, if one changes the profile of $\tilde{c}$ to be an elliptical Gaussian the integrals are no longer analytically tractable, but one can show that the radial cluster velocity will still be proportional to $\cos2\theta_0$, indicating that the discrepancy in transition angles is likely not due to anisotropy of the cluster shape.
The other assumption we make is that the region of anisotropy extends to infinity; however, in the simulations we maintained an anisotropic patterned region of radius $R = 0.5$.
If we increase $R$ and perform the simulations, the value of $\theta_c$ approaches $\theta_c \to \pi/4$ as shown in Fig.~\ref{fig:CriticalThetaDistance}.
To take into account the finite size of the anisotropic region in our analytical approximation would require finite bounds on the integrals involved in calculating $\mathbf{V}_c$, making them analytically intractable (the integrals are shown explicitly in the appendix).
We can perform the integration numerically, however, and in the inset of Fig.~\ref{fig:CriticalThetaDistance} we show that the numerically predicted critical angle is $\theta_c > \pi/4$ for finite $R/a$ and asymptotically approaches $\pi/4$ as $R/a$ increases.

\begin{figure}
    \centering
    \includegraphics[width = \columnwidth]{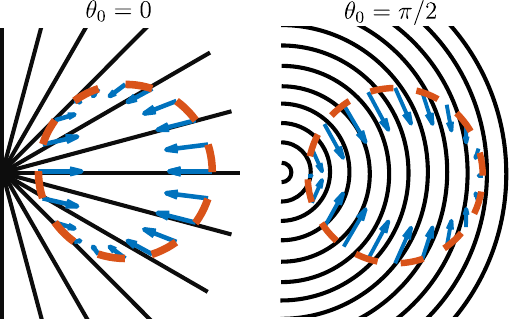}
    \caption{Approximate velocity field, Eq.~\eqref{eqn:ApproxVelocity}, plotted along a level set of a Gaussian cluster centered outside the center of a topological defect pattern for $\theta_0 = 0$ and $\theta_0 = \pi/2$.
    The asymmetry of the anisotropy across the profile leads to a net velocity of the cluster.}
    \label{fig:ActiveForceDefect}
\end{figure}

The qualitative success of our highly approximate model suggests a simple physical mechanism for the motion of clusters.
In general, asymmetry of the pattern across the cluster should induce motion because of the asymmetry in the fluid velocity field.
This is different than the above case shown in Fig.~\ref{fig:AnisotropicClustering}(a) for constant anisotropy since, in that case, there is no asymmetry across the cluster and the net velocity advecting the cluster is neutralized.
When the cluster is not at the center of the defect pattern, it is advected by a net velocity, as shown schematically in Fig.~\ref{fig:ActiveForceDefect}, where the velocity field is plotted along a circle centered outside the defect pattern for the cases of $\theta_0 = 0$ and $\theta_0 = \pi/2$.
For $\theta_0 = 0$, the velocity is more commensurate with the easy axis on the side of the cluster further from the defect center and more incommensurate with the easy axis on the other side, leading to a net motion inward.
It is precisely opposite for the case when $\theta_0 = \pi/2$, as shown in Fig.~\ref{fig:ActiveForceDefect}, so the cluster moves away from the center of the defect. %

\section{Patterning for Linear Motion} \label{sec:LinearPatterning}
\begin{figure}
    \centering
    \includegraphics[width = \columnwidth]{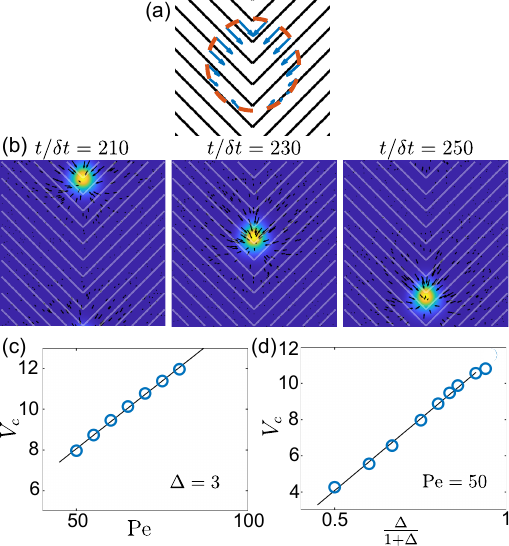}
    \caption{(a) Velocity field, Eq.~\eqref{eqn:ApproxVelocity}, on a level set of a Gaussian cluster in a chevron pattern of anisotropic friction.
    (b) Time snapshots of the local density (color) and velocity field (arrows) from a simulation with chevron patterned anisotropic friction.
    The overlaid lines give a visual indication of the anisotropic friction patterning.
    (c) Cluster velocity, as measured by the integral of the flux Eq.~\eqref{eqn:PeriodicFluxIntegral}, as a function of P{\'{e}}clet number for simulations with chevron anisotropic friction patterning and fixed $\Delta = 3$.
    The line shows a best fit to the data.
    (d) Cluster velocity as a function of $\Delta/(1+\Delta)$ for simulations with fixed $\text{Pe} = 50$.
    The line shows a best fit to the data for $\Delta \leq 10$.}
    \label{fig:ChevronMotion}
\end{figure}

We now show that our approximate model for cluster motion can be used to design a friction pattern that induces linear motion of a cluster.
Using the intuition gained from the motion of the cluster in topological defect patterns, we posit the following easy axis configuration:
\begin{equation} \label{eqn:ChevronPattern}
    \mathbf{\hat{e}}_{\parallel} = \begin{cases}
        \frac{1}{\sqrt{2}}\left[-\mathbf{\hat{x}} + \mathbf{\hat{y}}\right] & x < 0 \\
        \frac{1}{\sqrt{2}}\left[\mathbf{\hat{x}} + \mathbf{\hat{y}}\right] & x > 0.
    \end{cases}
\end{equation}
Equation \eqref{eqn:ChevronPattern} gives rise to a chevron pattern, which, if a cluster is centered at $x = 0$ should give rise to linear motion of the cluster based on the arguments of the previous section.
We show this schematically in Fig.~\ref{fig:ChevronMotion}(a).

To explicitly calculate the predicted cluster velocity, we start with the anisotropic part of the flux
\begin{equation} \label{eqn:ChevronJ}
    \mathbf{J}^\textrm{ani} = -\frac{2c_0^2}{\pi^2 a^6}\frac{\Delta \text{Pe}}{1 + \Delta}r\exp\left[-\frac{2r^2}{a^2}\right]\left[\tilde{\bm{\Gamma}}\mathbf{\hat{r}} \right]
\end{equation}
with
\begin{equation}
    \tilde{\bm{\Gamma}}\mathbf{\hat{r}} = \begin{cases}
         \frac{1}{2}\left[\cos\theta + \sin\theta\right]\left[\mathbf{\hat{x}} + \mathbf{\hat{y}}\right]& -\frac{\pi}{2} < \theta < \frac{\pi}{2} \\
        \frac{1}{2}\left[\cos\theta - \sin\theta\right]\left[\mathbf{\hat{x}}-\mathbf{\hat{y}}\right] & \frac{\pi}{2} < \theta < \frac{3\pi}{2}
    \end{cases}
\end{equation}
Integrating Eq.~\eqref{eqn:ChevronJ} gives a prediction for the cluster velocity:
\begin{equation} \label{eqn:ChevronV}
    \mathbf{V}_c = -\frac{c_0}{(2\pi)^{3/2} a^3}\frac{\Delta \text{Pe}}{1+\Delta} \mathbf{\hat{y}}.
\end{equation}
The direction of the predicted velocity agrees with the schematic analysis in Fig.~\ref{fig:ChevronMotion}(a), namely, the cluster is predicted to move in the $-y$ direction due to the top/bottom symmetry breaking.

Simulating the model with the chevron patterned anisotropic friction of Eq.~\eqref{eqn:ChevronPattern}, we find that clusters indeed move along the chevron pattern in the $-y$ direction, as shown in the time snapshots of Fig.~\ref{fig:ChevronMotion}.
A video of the simulation is also shown in Supplemental Movie 4 \cite{SuppNote25}.
The cluster clearly travels down, along the chevron pattern, as predicted by Eq.~\eqref{eqn:ChevronV}.
We also check the predicted scaling of Eq.~\eqref{eqn:ChevronV} with model parameters.
{To measure the cluster velocities in the simulations, we integrate the the flux as in Eq.~\eqref{eqn:CenterMassVelocity}.
We note that since we use periodic boundary conditions we must modify Eq.~\eqref{eqn:CenterMassVelocity} slightly so that
\begin{equation} \label{eqn:PeriodicFluxIntegral}
    \mathbf{V}_c = -\frac{1}{M}\oint_\partial\mathbf{r}\left(\bm{\hat{\nu}}\cdot\mathbf{J}\right)\,d\ell + \frac{1}{M}\int\mathbf{J}\, d\mathbf{r}
\end{equation}
where $\oint_\partial$ indicates a loop integral around the square boundary and $\bm{\hat{\nu}}$ is the outward normal of the boundary.}
In Fig.~\ref{fig:ChevronMotion}(c) we plot measured cluster velocities in simulations as a function of $\text{Pe}$ and find a linear scaling, as predicted by Eq.~\eqref{eqn:ChevronV}.
Additionally, in Fig.~\ref{fig:ChevronMotion}(d) we plot simulated cluster velocities as a function of $\Delta/(1+\Delta)$.
We find that the relationship is also linear, as predicted by Eq.~\eqref{eqn:ChevronV}, until $\Delta \gtrapprox 10$ at which point the cluster shape becomes highly asymmetric likely due to the different stability of clustering in different directions as explored in Sec.~\ref{sec:PhaseTransition}.
For $\Delta \gtrapprox 20$ we find that clusters are no longer stable and, instead, a one-dimensional band of density eventually forms along the center line.

\begin{figure}
    \centering
    \includegraphics[width = \columnwidth]{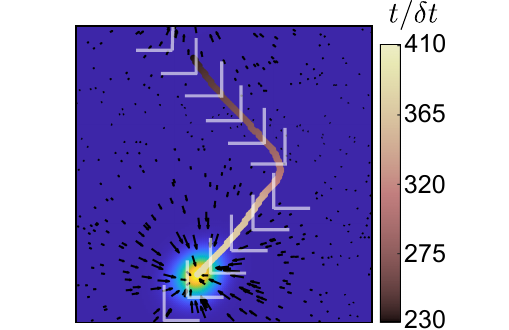}
    \caption{Simulated density field and velocity field at time step $t/\delta t= 410$ for a simulation with anisotropic friction pattern given by a zigzagging chevron pattern shown by the overlaid lines.
    The colored line is the trajectory of the cluster time steps $230 \leq t/\delta t \leq410$.
    }
    \label{fig:ZigZagTrajectory}
\end{figure}

Finally, we note that it is not necessary to cover the entire domain with the patterning to achieve linear cluster motion; rather, the width of the anisotropy region must just be large compared to the cluster size.
To show this, we perform a simulation with multiply directed chevrons so that the cluster changes direction.
In Fig.~\ref{fig:ZigZagTrajectory} we plot the configurations at time step $t/\delta t=410$ and plot the trajectory of the cluster between time steps $t/\delta t = 230$ and $t/\delta t = 410$.
The cluster moves along the anisotropic patterning, remaining inside the bounds of the anisotropic region (note that we set $\Delta = 0$ where there is no overlay in the figure).
Supplemental Movie 5 also provides a movie of the simulation \cite{SuppNote25}. %

\section{Discussion and Conclusion} \label{sec:Conclusions}

We have extended the classical Bois-J\"ulicher-Grill model of an active, compressible, isotropic fluid \cite{Bois11} to study two-dimensional fluids in the presence of anisotropic friction. 
As in the one-dimensional model, increasing active forces (increasing the P{\'{e}}clet number) leads to a transition between a homogeneous steady state and accumulation of material (clustering). Anisotropic friction may induce anisotropic clusters, or clustering in only one direction, depending on the anisotropy of the friction coefficients.
We further show that if the anisotropic friction is patterned in the fashion of an integer topological defect, stable clusters may form at the center of the defect, regardless of its geometry, as long as the configuration is initially radially symmetric.
For clusters that do not already sit at the center of the defect pattern, they may move either towards the center or away depending on the local asymmetry of the friction pattern, including persistent circular motion of the active cluster. We capture this transition with a simple theory balancing active forces with anisotropic friction. 
This understanding allows us to design substrates to create linear motion of clusters. 

Our results show a simple way to create directed migration in isotropic contractile active systems. The isotropy of the active fluid means the directionality is controlled by the symmetry of the environment. In contrast, active systems that have inherent symmetry breaking in the active stress, such as polar flocks or active nematics\cite{Vicsek95,marchetti13,doo18}, can develop macroscopic flows without an environmental asymmetry. It is possible that a combination of the environmental asymmetry we study here and clearer internal order in the active fluid could create new phenomena, e.g. using environmental asymmetry to select between spontaneously broken symmetries. 

Our work was initially motivated by experiments studying cells on anisotropic substrates\cite{Babakhanova20,turiv20,Endresen21,Kaiyrbekov23,Luo23,Zhao25}. While our model of an isotropic contractile fluid serves as a tractable initial model that can capture some features in pattern formation in fibroblasts \cite{Palmquist22}, additional order and symmetry may be present in the experimental systems. %
Cells on ridged anisotropic substrates can develop orientational order, with long axes of the cells parallel to the ridges -- ``contact guidance'' \cite{Petrie2009} -- suggesting they could be modeled as active nematic liquid crystals \cite{guillamat22,Blanch2021}.
However, other experiments reveal behaviors that do not conform to standard theories of active nematics. 
In particular, when cells are arranged in topological defect configurations, the geometry of the pattern does not qualitatively affect the motion of the cells\cite{Zhao25}, even though one expects {clear geometry-dependence in active nematics}\cite{doo18,schimming23b}.
Other results suggest that cellular monolayer systems may behave as extensile or contractile active nematics\cite{Luo23}, or even as passive nematics\cite{Duclos17,Murali25}, and other ordering features have also been observed \cite{Chiang2024,Eckert2023,Armengol-Collado2023-pk}. We also neglect proliferation of cells, which has been proposed to play a role in increased density in some of these experiments \cite{Kaiyrbekov23}. 

We note that single cells can also be guided by underlying asymmetries in the substrate, e.g. by asymmetric micropatterns or ratchet-shaped confinement \cite{Jiang2005,Mahmud09,Kushiro2010}. While these are also an example of active matter guided by substrate asymmetry, the driving factors are likely far more complicated than just the local friction. Predictions of migration on these patterns often require understanding of the protrusion dynamics of individual cells \cite{Caballero2014,Zadeh2024}.

Our results for the accumulation at the center of patterned defects are qualitatively similar to those of the experiments presented in Ref.~\cite{Zhao25}, with our Fig. \ref{fig:StableCenter} corresponding with their measurements of radial velocity in Fig. 3g and density in Fig. 1i.
Our contractile active fluid studied here is of course not a complete description of the measurements of \cite{Zhao25}, who see clear nematic order. However, we do capture the general pattern of radial velocities and how they depend on $\theta_0$. 
The model does not, however, predict the observation of \cite{Zhao25} that enhancement of density in the core is highest at $\theta_0 = \pi/3$. 
Our results show that some apparently characteristic features of active nematic cells can be captured with a simple isotropic model -- suggesting the need for care in interpreting the relevance of nematic order in a given system. 

More generally, an interesting interpretation of our results is that isotropic contractility, when paired with an anisotropic environment, may induce ostensibly extensile dynamics.
For example, when outside the center of the defect pattern we find that a cluster will move toward the center for $\theta_0 = 0$ (the ``aster'' configuration), but away from it for $\theta_0 = \pi/2$ (the ``target'' configuration).
In the language of nematic liquid crystals, the cluster moves along regions of positive splay and negative bend, which generate the direction of active force in an extensile active nematic.
Our results are consistent with another recent experiment on human dermal fibroblasts on a patterned liquid crystal elastomer substrate \cite{turiv20}.
They also explored both aster and target defect configurations, and found that clusters at $+1/2$ defects (which combine to form a $+1$ defect) migrated toward the core of the aster configuration, but away from the core of the target configuration.
Extensile active nematic dynamics was given as an explanation of these phenomena; however, the results of our model are also consistent with this behavior, and dermal fibroblasts are typically considered to be contractile \cite{Palmquist22}, not extensile.
This intriguing result requires further study as it may lead to a better understanding of why microscopically contractile systems may sometimes behave macroscopically as if they were extensile. %

\section*{Conflicts of interest}
There are no conflicts to declare.

\section*{Acknowledgements}
CDS gratefully acknowledges support from the Department of Physics and Astronomy at Johns Hopkins University through the William H. Miller III postdoctoral fellowship. BAC is supported by NIH R35 GM142847. We thank Dan Beller and Wei Wang for useful comments and a close reading of the draft. 

%\clearpage
\appendix

\section*{Appendix: Cluster Velocity in Defect Patterning Calculation} \label{app:Calculation}
Here we show the integrals leading to the prediction for the cluster velocity outside a patterned defect core, Eq.~\eqref{eqn:DefectClusterVel}, as well as the derivation of the integrals that are numerically computed for the inset of Fig.~\ref{fig:CriticalThetaDistance}.
For an infinitely patterned anisotropic region, we start by writing the integral of Eq.~\eqref{eqn:DefectAniCurrent}:
\begin{widetext}
%\begin{multline}
%\begin{equation}
\begin{multline}
    \mathbf{V}_c = -\frac{c_0}{2\pi^2 a^6}\frac{\Delta \text{Pe}}{1 + \Delta}e^{-\frac{2r_0^2}{a^2}}\int_0^{\infty}re^{-\frac{2r^2}{a^2}} 
    \int_0^{2\pi}e^{\frac{4r r_0}{a^2}\cos\theta} \\
    \left\{\left[r\cos\theta - r_0 + \left(r\cos\theta-r_0\cos2\theta\right)\cos2\theta_0  - (r\sin\theta + r_0\sin2\theta)\sin2\theta_0\right]\mathbf{\hat{x}} \right.   \\ +\left[r\sin\theta + \left(r\sin\theta - r_0\sin2\theta\right)\cos2\theta_0 \right.
    \left.\left.+\left(r\cos\theta - r_0\cos2\theta\right)\sin2\theta_0\right]\mathbf{\hat{y}}\right\}\,d\theta\,dr.
\end{multline}
%\end{equation}
The inner integral over $\theta$ is first computed yielding an expression in terms of modified Bessel functions of the first kind, $I_k$:

\begin{multline} 
\label{eqn:BesselIntegral}
    \mathbf{V}_c = -\frac{c_0}{\pi a^6}\frac{\Delta\text{Pe}}{1 + \Delta}e^{-\frac{2r_0^2}{a^2}}\int_0^{\infty}re^{-\frac{2r^2}{a^2}}
    \times\left\{\left[-r_0I_0\left(\frac{4r_0 r}{a^2}\right) + r(1+\cos2\theta_0)I_1\left(\frac{4r_0 r}{a^2}\right) - r_0\cos2\theta_0I_2\left(\frac{4 r_0 r}{a^2}\right)\right]\mathbf{\hat{x}} \right. \\
    \left. + \left[rI_1\left(\frac{4 r_0 r}{a^2}\right) - r_0I_2\left(\frac{4r_0 r}{a^2}\right)\right]\mathbf{\hat{y}}\right\}\,dr.
\end{multline}
\end{widetext}
To compute the integrals of the Bessel functions, we use a few standard integral identities \cite{gradshteyn2007} which give
\begin{align}
    \int_0^\infty r e^{-\frac{2r^2}{a^2}}I_0\left(\frac{4 r_0 r}{a^2}\right)\,dr &= \frac{a^2}{4}e^{\frac{2r_0^2}{a^2}} \\
    \int_0^\infty r^2 e^{-\frac{2r^2}{a^2}}I_1\left(\frac{4 r_0 r}{a^2}\right)\, dr &=\frac{r_0 a^2}{4}e^{\frac{2r_0^2}{a^2}}
    \end{align}
    and\begin{align}
    \int_0^{\infty}re^{-\frac{2r^2}{a^2}}I_2\left(\frac{4r_0 r}{a^2}\right)\, dr &= \frac{a^4}{8r_0^2}\left[1 + \left(\frac{2r_0^2}{a^2}-1\right)e^{\frac{2r_0^2}{a^2}}\right].
\end{align}
These identities yield Eq.~\eqref{eqn:DefectClusterVel} when used to compute the integrals in Eq.~\eqref{eqn:BesselIntegral}.

To derive the numerically computed integrals for the inset of Fig.~\ref{fig:CriticalThetaDistance}, where we assume that the friction anisotropy is only present in the region $r<R$ and outside that region $\Delta = 0$, we must also include the isotropic part of the flux, which we dropped above because it does not contribute to the net center of mass velocity.
Here, the isotropic part contributes to the center of mass velocity because the friction is inhomogeneous, i.e. $\Delta$ is a function of spatial coordinate and the integral of the first term in Eq. \ref{eq:flux_approximation} is no longer trivially zero.
For a region of anisotropy with spatial extent limited to a region of radius $R$, the approximate flux is
\begin{equation}
    \mathbf{J} = \begin{cases}
        \left[\left(\frac{\tilde{c}\text{Pe}}{1+\Delta}- 1\right)\mathbf{I} + \frac{\Delta \tilde{c}\text{Pe}}{1+\Delta}\bm{\tilde{\Gamma}}\right]\nabla \tilde{c} & r\leq R \\
        (\tilde{c}\text{Pe} -1)\nabla\tilde{c} & r> R
    \end{cases}
\end{equation}
hence, in addition to the integrals of the anisotropic part of the flux above, we must also account for the integrals on the isotropic part of the flux due to the difference in $\Delta$.
In what follows, we will only concern ourselves with the radial component of the center of mass velocity, since we are interested in understanding the discrepancy between radial transition angles in the simulations and theory.

The contribution to the center of mass velocity from the anisotropic part of the flux is given by Eq.~\eqref{eqn:BesselIntegral} with the upper limit of the integral changed to $R$.
The contribution from the isotropic part of the flux can be written as (ignoring contributions that are equal in the whole domain which integrate to zero) 
\begin{equation} \label{eqn:IsoCenterMassVel}
    V_{cr}^{iso} = -\frac{2c_0\text{Pe}}{\pi a^6}e^{-\frac{2r_0^2}{a^2}}\left[\frac{1}{1+\Delta}\mathcal{I}_0(0,R) + \mathcal{I}_0(R,\infty)\right]
\end{equation}
where $V_{cr}$ indicates the radial component of the velocity and
\begin{equation}
    \mathcal{I}_0(\alpha,\beta) = \int_\alpha^\beta r e^{-\frac{2r^2}{a^2}}\left[rI_1\left(\frac{4r_0 r}{a^2}\right) -r_0 I_0\left(\frac{4r_0 r}{a^2}\right)\right] \, dr.
\end{equation}
Since $\mathcal{I}_0(0,\infty) = 0$, we may rewrite Eq.~\eqref{eqn:IsoCenterMassVel} as
\begin{equation}
    V_{cr}^{iso} = -\frac{2 c_0}{\pi a^6}\frac{\Delta \text{Pe}}{1 + \Delta}e^{-\frac{2r_0^2}{a^2}}\mathcal{I}_0(R,\infty).
\end{equation}
Further, defining
\begin{equation}
    \mathcal{I}_1(\alpha,\beta) = \int_\alpha^\beta r e^{-\frac{2r^2}{a^2}}\left[rI_1\left(\frac{4r_0 r}{a^2}\right) -r_0 I_2\left(\frac{4r_0 r}{a^2}\right)\right] \, dr,
\end{equation}
we may write the full expression for $V_{cr}$:
\begin{equation}
    V_{cr} = -\frac{c_0}{\pi a^6}\frac{\Delta \text{Pe}}{1 + \Delta}e^{-\frac{2r_0^2}{a^2}}\left[\mathcal{I}_0(R,\infty) + \mathcal{I}_1(0,R)\cos2\theta_0\right].
\end{equation}
The critical patterning angle $\theta_c$ is defined by $V_{cr}(\theta_c) = 0$, hence we may finally write
\begin{equation} \label{eqn:CriticalAngle}
    \cos2\theta_c = \frac{\int_{R/a}^\infty u e^{-2u^2}\left[\frac{r_0}{a}I_0\left(\frac{4r_0}{a}u\right) - uI_1\left(\frac{4r_0}{a}u\right)\right]\,du}{\int_0^{R/a}u e^{-2u^2}\left[uI_1\left(\frac{4r_0}{a}u\right) - \frac{r_0}{a}I_2\left(\frac{4r_0}{a}u\right)\right]\,du}.
\end{equation}
Eq.~\eqref{eqn:CriticalAngle} is numerically integrated to produce the plot in the inset of Fig.~\ref{fig:CriticalThetaDistance} for $r_0/a = 1$.

\bibliography{LC}

\end{document}

\title{Supplemental Material for ``Guiding isotropic active fluids with anisotropic friction''}

\author{Cody D. Schimming}
\affiliation{William H. Miller III Department of Physics and Astronomy, Johns Hopkins University, Baltimore, Maryland 21218, USA.}

\author{Brian A. Camley}
\affiliation{William H. Miller III Department of Physics and Astronomy, Johns Hopkins University, Baltimore, Maryland 21218, USA.}
\affiliation{Thomas C. Jenkins Department of Biophysics, Johns Hopkins University, Baltimore, Maryland 21218, USA.}

\maketitle

\section*{Supplemental Movie Captions}
\begin{itemize}
    \item Supplemental Movie 1: Simulated density field (color) and velocity field (arrows) for topological defect friction anisotropy patterning with $\theta_0 = \pi/2$ and an azimuthally symmetric, annular density initial condition.
    \item Supplemental Movie 2: Simulated density field and velocity field for topological defect friction anisotropy patterning with $\theta_0 = 0$ and an off-center Gaussian density initial condition.
    \item Supplemental Movie 3: Simulated density field and velocity field for topological defect friction anisotropy patterning with $\theta_0 = \pi/3$ and an off-center Gaussian density initial condition.
    \item Supplemental Movie 4: Simulated density field and velocity field for a linear chevron friction anisotropy patterning.
    \item Supplemental Movie 5: Simulated density field and velocity field for a zigzag chevron friction anisotropy patterning.
\end{itemize}

\renewcommand{\thefigure}{S1}
\begin{figure*}[h]
    \centering
    \includegraphics[width = \textwidth]{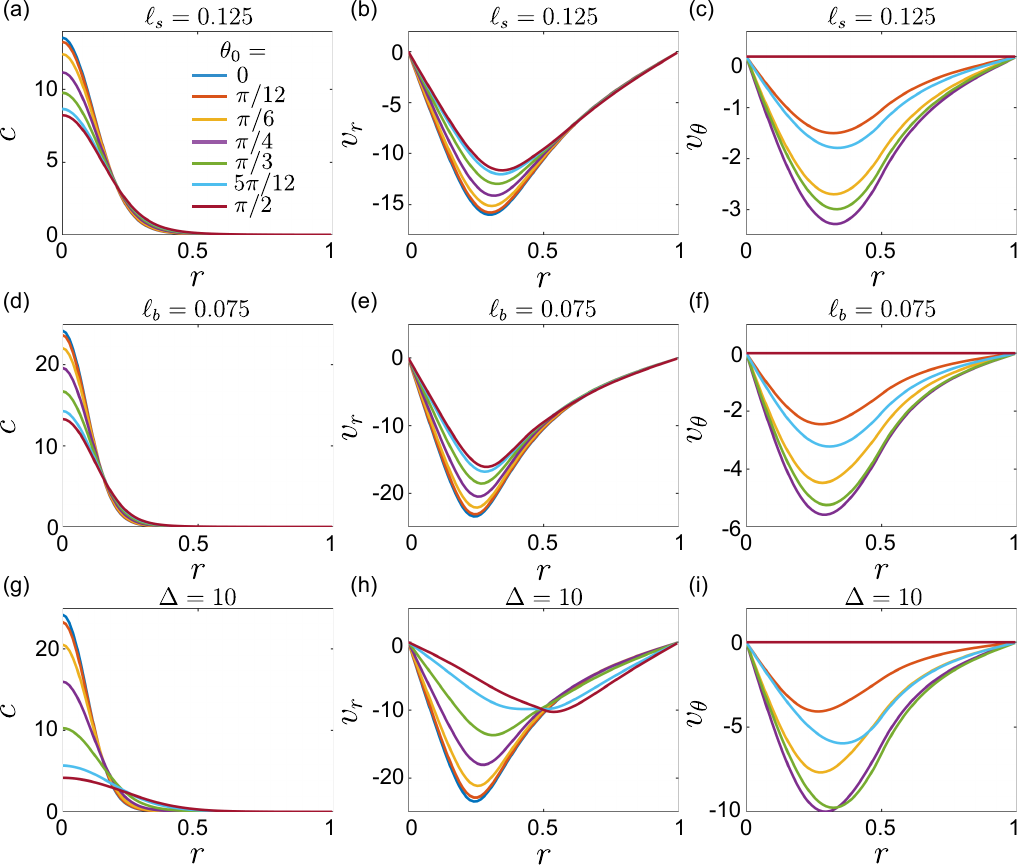}
    \caption{Radial cuts of simulated stable clusters at the center of topological defect patterned anisotropic friction regions for a variety of patterning angles $\theta_0$.
    (a,d,g) Density as a function of distance, $c(r)$.
    (b,e,h) Radial component of the velocity field as a function of distance $v_r(r)$.
    (c,f,i) Azimuthal component of the velocity field as a function of distance $v_\theta(r)$.
    For all simulations, parameters are given in Table 1 of the main text except for (a--c) $\ell_s = 0.125$, (d--f) $\ell_b = 0.075$, (g--i) $\Delta = 10$.}
    \label{fig:ParameterVariation}
\end{figure*}

\bibliography{LC}